\documentclass[%
 aip,
 amsmath,amssymb,
 preprint,%
]{revtex4-1}

\usepackage{graphicx}
\usepackage{bm}
\usepackage{xcolor}
\usepackage{soul}
\usepackage{setspace}

\usepackage[T1]{fontenc}
\usepackage{mathptmx}
\begin{document}

\title[]{Characterizing signal encoding and transmission in class I and class II neurons via ordinal time-series analysis}

\author{C. Estarellas}
\email{cristian@ifisc.uib-csic.es}
 \affiliation{Institut de F\'{i}sica Interdisciplinar i Sistemes Complexes (IFISC, UIB-CSIC), Campus Universitat de les Illes Balears E-07122, Palma de Mallorca, Spain}
\author{M. Masoliver}%
 \email{maria.masoliver@gmail.com}
\affiliation{Departament de F\'{i}sica, Universitat Polit\`{e}cnica de Catalunya, Terrassa 08222, Spain
}%

\author{C. Masoller}
\affiliation{Departament de F\'{i}sica, Universitat Polit\`{e}cnica de Catalunya, Terrassa 08222, Spain
}%
\author{Claudio R. Mirasso}
\affiliation{Institut de F\'{i}sica Interdisciplinar i Sistemes Complexes (IFISC, UIB-CSIC), Campus Universitat de les Illes Balears E-07122, Palma de Mallorca, Spain}
\date{\today}

\begin{abstract}
{\setstretch{1.6}

Neurons encode and transmit information in spike sequences. However, despite the effort devoted to quantify their information content, little progress has been made in this regard.
Here we use a nonlinear method of time-series analysis (known as ordinal analysis) to compare the statistics of spike sequences generated by applying an input signal to the neuronal model of Morris-Lecar.
In particular we consider two different regimes for the neurons which lead to two classes of excitability: class I, where the frequency-current curve is continuous and class II, where the frequency-current curve is  discontinuous.
 By applying ordinal analysis to sequences of inter-spike-intervals (ISIs) our goals are (1) to investigate if different neuron types can generate spike sequences which have similar symbolic properties; 
 (2) to get deeper understanding on the effects that electrical (diffusive) and excitatory chemical (i.e., excitatory synapse) couplings have; and (3) to compare, when a small--amplitude periodic signal is applied to one of the neurons, how the signal features (amplitude and frequency) are encoded and transmitted in the generated ISI sequences for both class I and class II type neurons and electrical or chemical couplings. We find that depending on the frequency, specific combinations of neuron/class and coupling-type allow a more effective encoding, or a more effective transmission of the signal. }
\end{abstract}
\maketitle

\begin{quote}
 \textbf{Sensory neurons detect, encode and transmit information of external temporal stimuli (input signals such as visual, auditory, olfactory, etc.) in sequences of spikes, also known as action potentials. 
Despite decades of effort to understand how information is processed, the underlying mechanisms of neuronal encoding are still not fully understood. 
Different coding mechanisms have been proposed in the literature, which can be more or less effective depending on the level of environmental noise, the level of the external signal, and its frequency. 
Here we focus on the encoding of a small-amplitude periodic signal. 
We use a well-known neuron model to investigate the role of the excitability class of the neurons (either class I or class II) and of the type of coupling (electrical or chemical) to a second neuron that does not perceive the external signal. 
We find that the neuron can encode the signal in the form of preferred (more expressed) temporal spike patterns, regardless of the class of neuron and of the type of coupling. 
On the contrary, depending on the signal frequency, specific combinations of neuron-class and coupling-type allow a more effective encoding, or a more effective transmission of the signal. }
\end{quote}

\section{Introduction \label{Intro}} 

The neuronal mechanisms used for encoding and transmitting information have been extensively studied \cite{thorpe_2001, nature_2002, nature_2003, hidden_2004, coombes_2010}, yet they are still unclear. 
The neural code can use the activity of single neurons \cite{QUI05} or the activity of a population of neurons, and can be based on the timing of the spikes (temporal coding), on the frequency of the spikes (rate coding), on the phase of spike firings, on the correlations between the spikes, etc. \cite{STE05}. 
As sensory neurons respond to a wide variety of external stimuli (motor, visual, olfactory, auditory, etc.), different mechanisms are likely used, depending on the type of signal, on the signal features, and on the signal to noise ratio.

In a recent work Reinoso et al.\cite{REI16} used the well-known FitzHugh-Nagumo model \cite{NAGU_55,NAGU_61} to investigate how a single noisy neuron encodes a periodic signal. 
The signal was weak enough to be subthreshold (of small amplitude and/or of slow period \cite{LON98}) and in the absence of noise, it did not generate spiking activity. 
It was shown that the period and the amplitude of the weak periodic signal could be encoded in the relative timing of the spikes (inter-spike intervals) that were fired due to the presence of noise. 
The study used a non-linear method of time series analysis known as \emph{ordinal analysis} \cite{pre_2009, amigo, review}. 
Ordinal analysis is a technique which transforms a particular sequence (here the inter-spike interval sequence of the spike-train of the neuron) into a sequence of symbols known as ordinal patterns; probabilities for each symbol (ordinal patterns probabilities) are then computed. 
Ordinal pattern probabilities determine if some symbols are more or less expressed revealing (or not) some degree of order on the spike-train of the neuron. 
Reinoso and collaborators found that the ordinal probabilities depended on both the period and the amplitude of the applied signal. 
In a follow up study, Masoliver and Masoller \cite{MAS18} demonstrated the robustness of this encoding mechanism when a second neuron, which does not directly perceive the signal, is coupled to the neuron that received the signal. 
It was demonstrated that the encoding mechanism is robust, as ordinal probabilities of the spike train of the neuron that perceives the signal still carry information about the features of the signal (amplitude and period). 

Both previous studies were centered on the encoding of the signal, but not on its transmission. 
In this paper we address how a signal is not only encoded but also transmitted from the neuron that receives it, to a second neuron that does not directly perceive the signal. 
We consider the Morris-Lecar model \cite{ML} that allows us to investigate two type of neurons with different excitability class: class I and class II. Spikes generated by class I neurons depend on the strength of the applied current but can be generated with arbitrary low frequency whereas spikes generated by class II neurons hardly depend on the frequency (just a certain frequency band is allowed) but not on the amplitude \cite{IZH10}.

We compare the spike trains of the two neurons (the one that receives the signal and the one that does not directly receive it), in order to determine whether the signal information (amplitude and period) is present in the spike train of the second neuron. 
In particular, we aim at addressing the following questions (1) are the encoding and transmission mechanisms independent of the class of neuron, i.e., class I excitability vs class II excitability neurons (for example, cortical fast spiking vs regular spiking neurons)? (2) is the encoding mechanism modified  by the type of synapse (electrical or chemical) between neurons? and (3) does the transmission depend on the synapse?

The outline of this paper is as follows. In section \ref{model} we introduce the model and describe the behavior of two-coupled Morris-Lecar neurons.
We also introduce the different coupling schemes and give some insights into the synaptic currents. 
Section \ref{analysis} focuses on the formalism of ordinal analysis. 
In section \ref{result} we present the numerical results regarding (A) the robustness of symbolic analysis (B) the encoding of the signal focusing on different classes of neurons (class I and class II) and the types of synapse (chemical and electrical) and (C) the transmission of the signal depending on the class of neuron and type of synapse. 
In section \ref{conclu} we present a summary and conclusions.

\section{\label{model}Model}

\subsection{\label{sec:neuronal} Morris-Lecar model}
The two-dimensional model considered by Prescott \cite{Pres}, as a variation of the Morris-Lecar model \cite{ML}, is described by:

\begin{align}
\label{dvdt}
C\frac{dV}{dt}  =&I_{stim}-g_{fast}m_{\infty}(V)(V-E_{Na})-g_{slow}w(V-E_{k})\nonumber\\
 &  -g_{leak}(V-E_{leak}),\nonumber \\
  \frac{dW}{dt}=& \phi_w\frac{w_{\infty}(V)-w}{\tau_{w}(V)}.
\end{align}

The voltage ($V$) of the membrane potential is the fast variable, and the activation of the potassium channel ($W$) is the slower variable. 
The parameters $E_k$, $E_{leak}$ and $E_{Na}$ are the equilibrium potential, and $g_{fast}$, $g_{slow}$ and $g_{leak}$ represents the maximal conductance for $Na^+$, $K^+$ and leak currents. 
Equations (\ref{state}) and (\ref{time}) characterize the steady-state of the activation function of the $Na^+$ and $K^+$ channel and the time constant of the last one.

\begin{eqnarray}
\label{state}
& m_{\infty}(V)=0.5\left[1+\tanh\left(\frac{V-\beta_m}{\gamma_m}\right)\right],\nonumber\\
& w_{\infty}(V)=0.5\left[1+\tanh\left(\frac{V-\beta_w}{\gamma_w}\right)\right],
\end{eqnarray}

\begin{equation}
\label{time}
\begin{split}
& \tau_{w}(V)=\frac{1}{\cosh\left(\frac{V-\beta_w}{2\gamma_w}\right)}.\\
\end{split}
\end{equation}

The main parameters of the model are the same as those used in the Prescott's study \cite{Pres}:\\

$E_{Na}=50$ mV, $E_{K}=-100$ mV, $E_{leak}=-70$ mV, $g_{fast}=20 $mS/cm$^2$, $g_{slow}=20$ mS/cm$^2$, $g_{leak}=2$ mS/cm$^2$, $\phi_w=0.15$, $C=2$ $\mu$F/cm$^2$, $\gamma_m=18$ mV.\\

The Morris-Lecar model is known to generate different dynamical regimes depending on its parameters \cite{Bi1,Liu}. 
We fixed the values of $\beta_w=-10$ mV and $\gamma_w=13$ mV and used two values of $\beta_m$ to obtain the two classes of excitability. 
In particular, for a class I neuron $\beta_m=-12$ mV and for a class 2 $\beta_m=0$ mV (see Appendix \ref{Appendix}).

The stimulating current ($I_{stim}$) is composed of three contributions. 
The first one is the synaptic current, representing a chemical or electrical synapsis (see section \ref{Scurrents}) projected by the other neuron. 
The second one is the Poisson noise; this noise accounts for the contribution of external neurons that are not included in the model. 
The distribution of action potentials that arrive at the neuron follows the Poisson distribution, and each event produces a chemical excitatory synapse, whose intensity is regulated by the synaptic conductance ($g_{po}$). 
This contribution is assumed to be suprathreshold because it generates spontaneous spikes in the receiver neuron. 
The last input is a sinusoidal modulation with zero mean value that we apply in one of the neurons. This current is assumed to be subthreshold, therefore, without noise, it cannot generate spikes in the target neuron.

\begin{figure}[!ht]
\centering
\includegraphics[width=0.6\columnwidth]{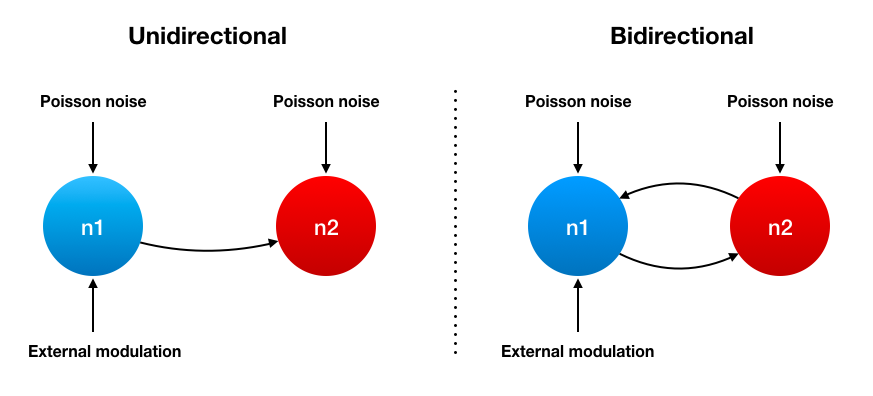}
\caption{Scheme of the unidirectional (a) and bidirectional (b) models. For both models neurons have Poisson noise and the neuron 1 (blue) receives an external modulation; and neuron 2 (red) is coupled to neuron 1. For the unidirectional model neuron 1 is not coupled to neuron 2, yet for the bidirectional model neuron 1 is coupled to neuron 2.}
\label{Scheme}
\end{figure}

The possible couplings between the neurons is schematically represented in Fig. \ref{Scheme}. Fig. \ref{Scheme}(a) accounts for the unidirectional case: the neuron that receives the signal (named as neuron 1) is connected to neuron 2 (the unmodulated one) but neuron 2 is not connected to neuron 1. Fig. \ref{Scheme}(b) accounts for the bidirectional case: both neurons are connected to each other. We are going to study the two types of possible connections: chemical and electrical synapses. For this last case, due to the diffusive effect of the gap junction the plausible situation is the bidirectional case. However, to compare the two classes of excitability in this study the electrical coupling is used for the unidirectional model as well. 

\subsection{ \label{Scurrents}Synaptic Currents}
One key point of this study is the discussion of the effect that the two potential synaptic mechanisms (chemical and electrical) have on the encoding and transmission of the signal.
In this study, only excitatory synaptic current are considered. These currents are mediated by AMPA receptors and are described by:
\begin{equation}
\label{Ix}
I_{A} = -g_{A}r_{A}(V-E_A).
\end{equation}

The parameter $g_{A}$ is the maximal conductance and is fitted to obtain a phase locking 1-1 regime between the coupled neurons. 
The reversal potential is $E_A=0$ mV, and $r_A$ represents the probability of the presynaptic neurotransmitters to be released and caught by the postsynaptic receptors. 
Its dynamics is approximated by: 

\begin{equation}
\label{drdt}
  \tau_A\frac{dr_{A}}{dt}=-r_{A} + \sum_{k\epsilon \Omega} \delta(t-t_k),\\
\end{equation}

\noindent where $\Omega$ is the group of presynaptic neurons that generate an action potential at time $t_k$ and $\tau_A=5.6$ ms is the decay time of the chemical synapse.\\
The electrical synapse provides a diffusive coupling  between the postsynaptic ($i$) and the presynaptic ($j$) neurons expressed by:

\begin{equation}
\label{gap}
I_{i} = -g_{gap}(V_i-V_j),
\end{equation}

\subsection{Parameter values}
\label{param_values}
\begin{figure}[!ht]
\centering
\includegraphics[width=0.6\columnwidth]{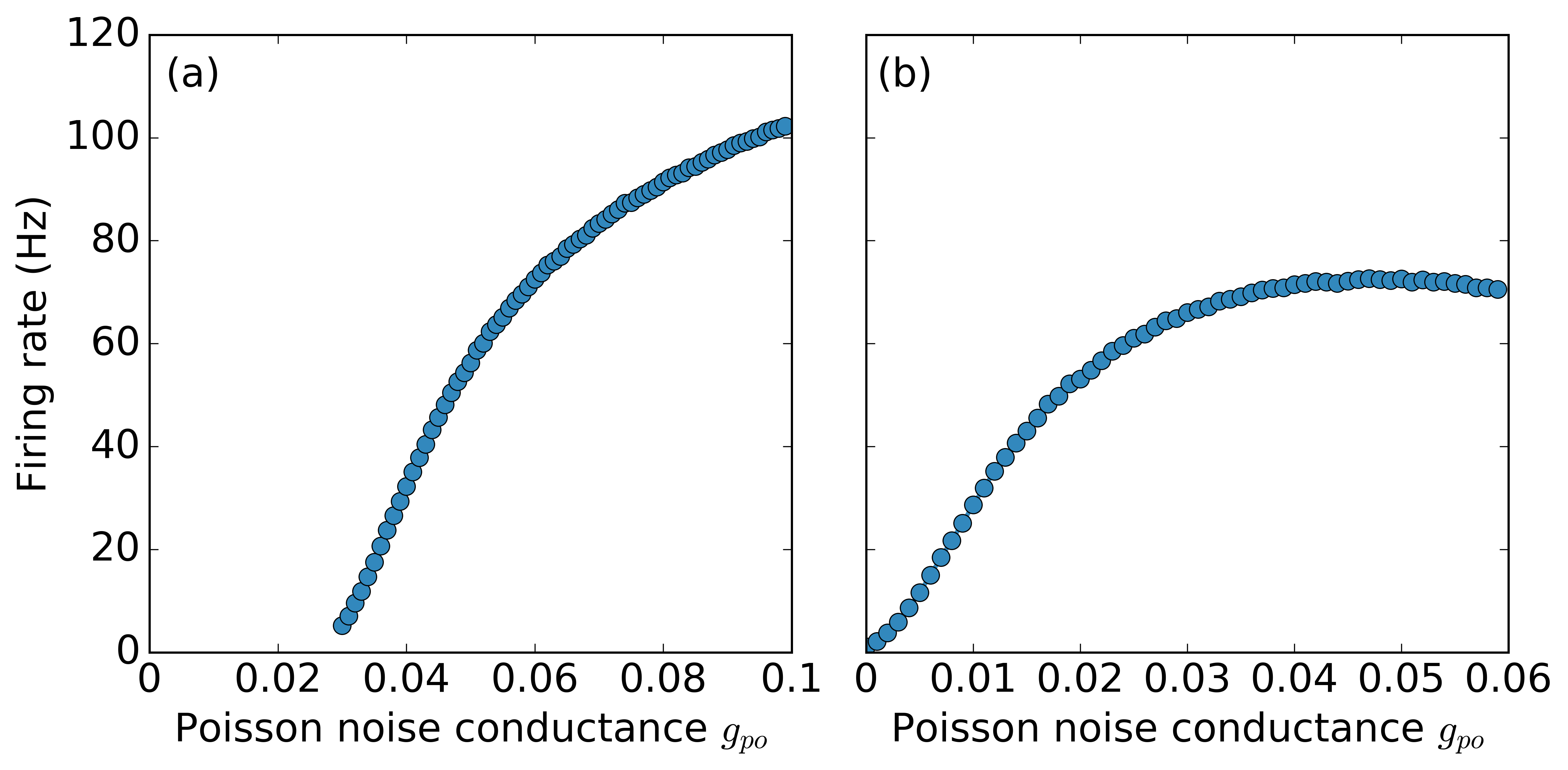}
\caption{Firing rate of neuron 1 uncoupled and without external modulation as a function of the synaptic Poisson noise conductance $g_{po}$ for class I and class II types, (a) and (b) respectively.}
\label{Fig1}
\end{figure}

In order to compare excitable class I and class II neuron types we first analyze how the firing rate of an individual neuron varies with the external noise (Fig. \ref{Fig1}a and \ref{Fig1}b). 
Whenever we fix the noise intensities, we will use those values that lead to the same firing rate; for class I, $g_{p_I} = 0.03$ mS/cm$^2$ and for class II $g_{p_{II}} = 0.19$ mS/cm$^2$, which lead to a firing rate of $f_r \approx 8.6$ Hz. 
The external modulation amplitudes for class I and class II are normalized by $g_{p_I}$ and $g_{p_{II}}$, as shown in equation (\ref{norm_A}) so that we have a unique modulation amplitude for both classes, $A_0$. 
Whenever it is fixed, its value will be $A_0 = 250$ mV~cm$^2$/mS. 
We will also vary the modulation amplitude for both neuronal classes from 0 to 400 mV~cm$^2$/mS; it is important to remark that for class I neurons the modulation is sub-threshold for all the regime, while for class II the signal is supra-threshold for $A_0 \geq 325$ mV~cm$^2$/mS. 

\begin{eqnarray}
\frac{A_{m_I}}{g_{p_I}} = \frac{A_{m_{II}}}{g_{p_{II}}} = A_0
\label{norm_A}
\end{eqnarray}
Similarly, as shown in equation (\ref{norm_cond}), the electrical ($g_{gap_{I,II}}$) and chemical ($g_{AMPA_{I,II}}$) conductances are also normalized for both class I and class II neurons, $g_{gap_{norm}}$ and $g_{AMPA_{norm}}$. 
Whenever the conductance values are fixed, we take $g_{gap_{norm}} = 10$ (for both unidirectional and bidirectional cases), $g_{AMPA_{norm}} = 16.6$ (unidirectional) and $g_{AMPA_{norm}} = 10$ (bidirectional).

\begin{eqnarray}
\frac{g_{gap_I}}{g_{p_I}} = \frac{g_{gap_{II}}}{g_{p_{II}}} = g_{gap_{norm}}\nonumber\\
\frac{g_{AMPA_I}}{g_{p_I}} = \frac{g_{AMPA_{II}}}{g_{p_{II}}} = g_{AMPA_{norm}}
\label{norm_cond}
\end{eqnarray}

\section{\label{analysis}Symbolic time series analysis}

Ordinal analysis is a symbolic technique which has been used to detect correlations in time-series data which were not detected using regular lineal tools of time-series analysis \cite{REI16, BAN02, REI16b, ROS09}. 
Here, ordinal analysis is applied to the sequence of inter-spike intervals  $\{I_1, \dots, I_i, \dots I_N \}$,  and a sequence of symbols, named ordinal patterns is obtained. 
Ordinal patterns depend on the temporal order between $L$ consecutive inter-spike intervals.  
For each inter-spike interval the subsequents $L-1$ are compared. 
The total number of possible symbols is then equal to $L!$.
 If $L = 2$ there are $2!$ (two) possible order relations (1) $I_1 < I_2$ and (2) $I_1 > I_2$ (in the case that $I_1 = I_2$ a small random noise is added to decide which one is larger than the other). 
All through the paper we consider $L = 3$ (it allows to investigate the order relation among 3 inter-spike intervals) which gives $3! = 6$ possible symbols, six ordinal patterns. 
Once the sequence of ordinal patterns is computed (using the function \texttt{perm} \underline{  } \texttt{indices}  defined in \cite{PAR12}), ordinal patterns probabilities and permutation entropy are computed. 
Ordinal pattern probabilities are estimated as $p_i = N_i/M$ where $N_i$ denotes the number of times the i-th pattern appears in the sequence, and $M$ denotes the total number of patterns. 
If the patterns are equiprobable one can infer that there are no preferred order relations in the timing of the spikes. 
On the contrary, the presence of frequent (or infrequent) patterns will result into a non-uniform distribution of the ordinal patterns. 
A binomial test is used to analyze the significance of preferred and infrequent patterns: if all the ordinal probabilities are within the interval $[p - 3\sigma_p, p + 3\sigma_p]$ (with $p = 1/L!$ and $\sigma_p = \sqrt{p(1-p)/M}$), the probabilities are consistent with the uniform distribution. Otherwise, there are important deviations which reveal the presence of more and less expressed patterns. In order to quantify at once these deviations we compute the permutation entropy $H$, defined as 

\begin{equation}
H =\frac{\sum_i^{L!} p_i \log p_i}{\log L!} 
\end{equation}

which ranges between 0 (i.e., regular and deterministic behavior) and 1 (i.e., completely noisy and random behavior). Already small deviations from $H = 1$ can be used to identify deviations from a fully noisy behavior. It is associated to the array of ordinal probabilities $p_i$ with $i = {1, \dots L!}$ \cite{REI16,CAO04,ROS07,ARA16}.

\section{\label{result}Results}
\subsection{\label{robust} Robustness of the signal encoding via ordinal patterns}
\begin{figure}[!ht]
\centering
\includegraphics[width=0.6\columnwidth]{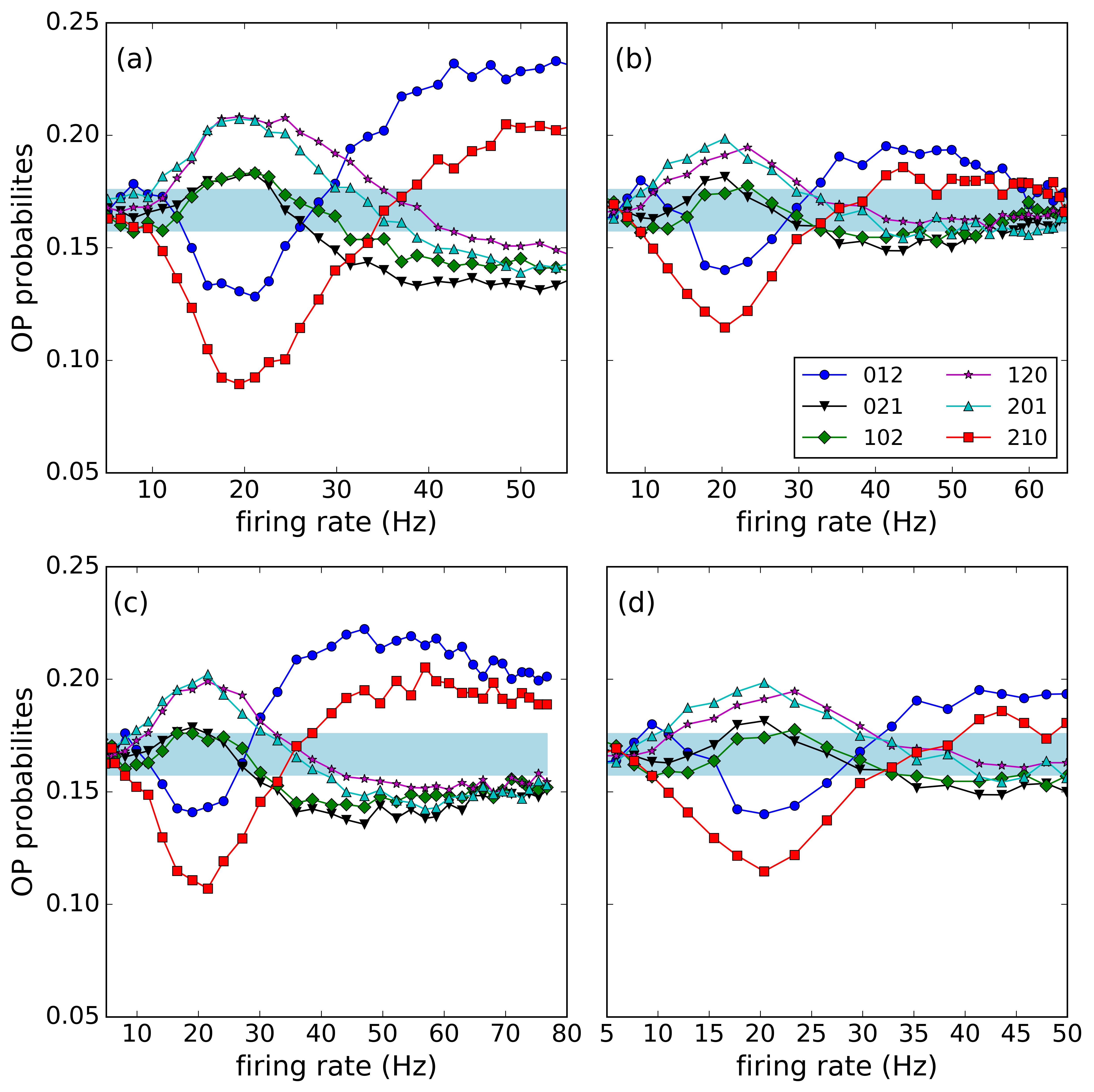}
\caption{Ordinal patterns probabilities as a function of the spike rate for neuron 1 for the bidirectional coupling case. Panels (a) and (b) correspond to class I and class II, respectively, with chemical coupling $g_{{AMPA}_{norm}} = 10$. Panels (c) and (d) correspond to class I and class II, respectively, with electrical coupling $g_{{gap}_{norm}} = 10$. Other parameters: $f = 10Hz$ and $A_0 = 250$ mVcm$^2$/mS. Noise amplitude was within the range $0.1$ mS/cm$^2$ $< g_{po} <$ $0.5$ mS/cm$^2$.}
\label{robust1}
\end{figure}

In a recent study \cite{MAS18}, the encoding of a weak  external signal by a neuron (modeled with the FitzHugh-Nagumo model) mutually coupled to a second neuron (which did not directly receive the signal) was studied. 
There it was shown that the encoding mechanism was robust to the coupling: the modulated neuron carried, under certain circumstances, information about the signal. 
Interestingly, ordinal patterns probabilities exhibited a resonant effect when the external frequency of the signal was twice the spike rate of the neuron.
In Figure \ref{robust1} we show how the encoding mechanism is robust to different models, coupling types and neuron excitability classes. 
As in \cite{MAS18} we observe, for the two variations of the Morris-Lecar model, the resonant effect for both types of coupling (chemical and electrical) and for the two type of neurons (class I and class II). 
Figure \ref{robust1} displays ordinal patterns probabilities as a function of the firing rate. 
In the four panels, we observe that P(210) has a minimum when the firing rate is close to 20 Hz, twice the applied modulation frequency. 

\subsection{\label{enco}Quantifying information encoding}
\begin{figure}[!ht]
\centering
\includegraphics[width=0.6\columnwidth]{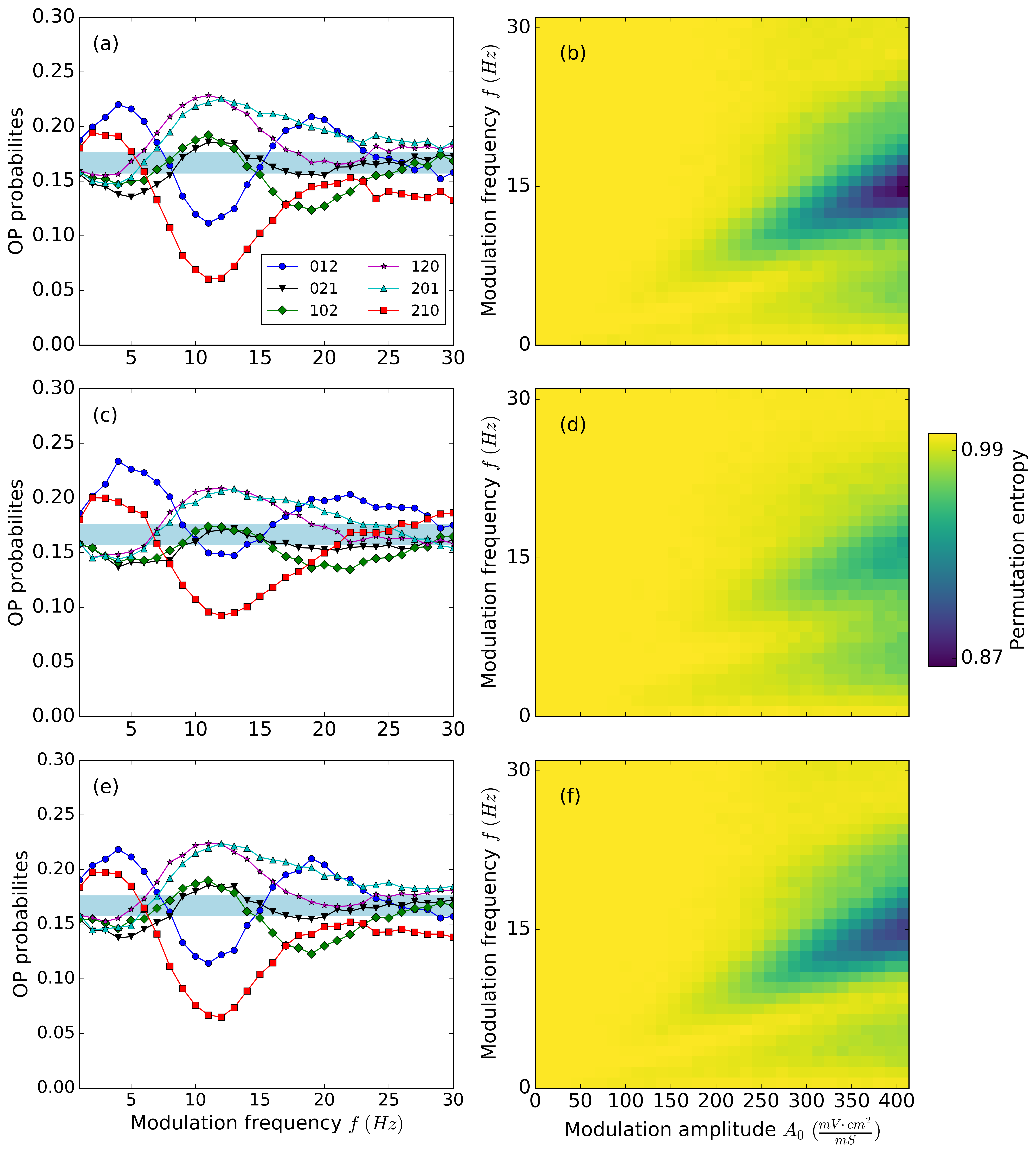}
\caption{Permutation entropy in color code as a function of the modulation frequency and the normalized modulation amplitude  for class I and for the unidirectional model, panels (b), (d) and (f); OP probabilities vs. the modulation frequency for $A_{0}=250$ mVcm$^2$/mS, panels (a), (c) and (e). Panel (a) and (b) correspond to neuron 1; panel (c) and (d) to neuron 2 with chemical coupling $g_{AMPA_{norm}}$; panel (e) and (f) to neuron 2 with electrical coupling $g_{{gap}_{norm}}$. For specific values see section \ref{param_values}.}
\label{T1_1}
\end{figure}

\begin{figure}[!ht]
\centering
\includegraphics[width=0.6\columnwidth]{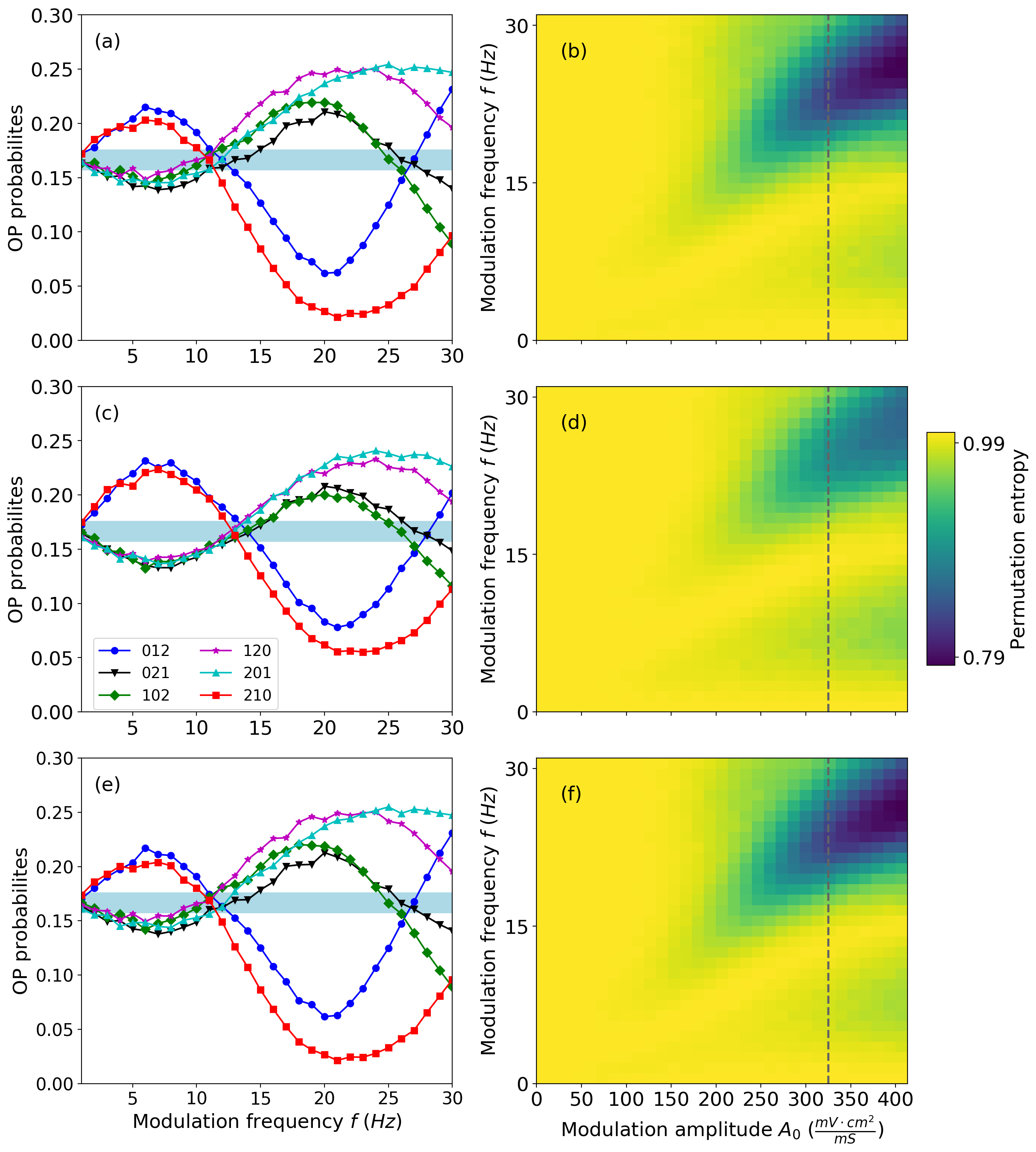}
\caption{Permutation entropy in color code as a function of the modulation frequency and the normalized modulation amplitude  for class II and for the unidirectional model, panels (b), (d) and (f), dotted grey line marks when the modulation becomes supra-threshold ($A_0 \geqslant 325$ mVcm$^2$/mS); OP probabilities vs. the modulation frequency for $A_{0}=250$ mVcm$^2$/mS, panels (a), (c) and (e). Panel (a) and (b) correspond to neuron 1; panel (c) and (d) to neuron 2 with chemical coupling $g_{AMPA_{norm}}$; panel (e) and (f) to neuron 2 with electrical coupling $g_{{gap}_{norm}}$. For specific values see section \ref{param_values}.}
\label{T1_2}
\end{figure}

The efficacy of the encoding can be understood as the number of possible spiking patterns that a neuron can generate due to different inputs. 
The quality of the encoding can be defined as the robustness of a spiking pattern generated by a particular input. 
In this sense, the variation of the maxima OP probabilities generated by the neuron, for different frequencies of the external modulations, accounts for the efficacy and quality of the neural encoding.
In this sense, when the neurons are unidirectionally coupled [Figure \ref{Scheme}(a)], a class I excitable neuron exhibits two maxima between $0-10$ and $17-25$ Hz and a minimum between $10-17$ Hz for the 012 pattern [represented with blue color in Figure \ref{T1_1}(a), (b) and (c)].  
Thus, the neuron can encode three ranges of external modulation frequencies. 
On the contrary, in the excitability class II case, the neuron can only encode two ranges of frequencies [Figure \ref{T1_2}(a), (b) and (c)]. The encoding is also better in class I since the OP probabilities have more significant differences between maxima and minima than in the case of class II. 

Another difference between the two excitability classes is the frequency range of the external signal that the modulated neuron can identify. 
One way to characterize this problem is to use permutation entropy (see section III).
In Figure \ref{T1_1}(b), (d) and (f) permutation entropy is plotted as a function of the modulation frequency and the modulation amplitude for class I neurons. It can be seen that the permutation entropy decreases for slow--frequency signals. ($f\lesssim 15$ $ Hz$).
On the contrary, in class II neurons it decreases for high--frequency signals  ($f\gtrsim20$ $Hz$), as can be seen in Figure \ref{T1_2}(b), resulting in a more orderly pattern. 
However, both class of neurons start to encode information for a similar minimum amplitude of the external modulation ($A_{0}\approx 200$ mVcm$^2$/mS).
Likewise, the permutation entropy decreases gradually (OP probabilities leave within the blue region) as the amplitude increases.
A closer look at the permutation entropies and the OP probabilities in Figure \ref{T1_1} and Figure \ref{T1_2} reveals that when the OP probabilities are larger, the value of the permutation entropy is smaller (for large enough modulation amplitude) reflecting that the encoding efficiency is larger. 
This difference is better seen for class II neurons where the difference between OP probabilities are larger, in particular at high frequencies, as compared to class I neurons.

\begin{figure}[!ht]
\centering
\includegraphics[width=0.6\columnwidth]{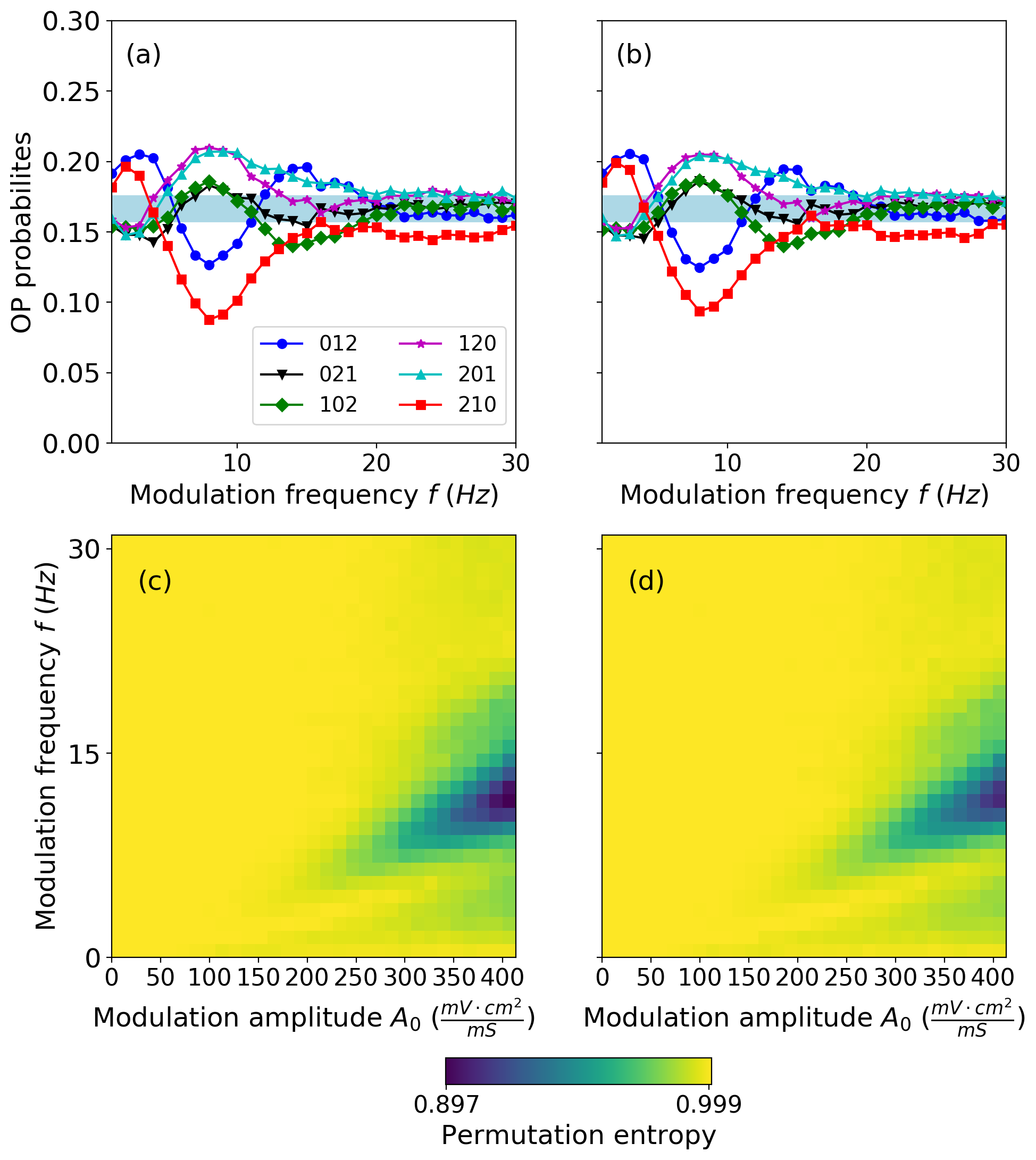}
\caption{OP probabilities as a function of the modulation frequency, for neuron 1 (panel a) and neuron 2 (panel b), for $A_{0} = 250$ mVcm$^2$/mS. Permutation entropy, in color map, as a function of the frequency and the normalized amplitude of the external modulation for neuron 1 (panel c)  and neuron 2 (panel d). All panels correspond to class I and to bidirectional electrical coupling with $g_{{gap}_{norm}}$. For specific values see section \ref{param_values}.}
\label{E1}
\end{figure}

In the bidirectional case, Figure \ref{Scheme}(b), the neuron that is subject to the external modulation is affected by the feedback from the second neuron. 
It is well known that feedback can considerably affect the dynamics of the neurons and so the encoding and transmission capabilities. 
We find that the encoding capability of the neuron strongly depends on the kind of coupling of the system. 
For the electrical coupling, and while we keep the 1:1 locking regime, for class I the response of the bidirectional system (Figure \ref{E1}) does not have significant differences when compared to the unidirectional case, Figure \ref{T1_1}(c). This is reflected in both the OP probabilities and the permutation entropy. Due to the diffusive effect of the coupling, both neurons synchronize achieving the same dynamics. 
For class II (Fig. \ref{Eel}), we observe as well that both neurons synchronize: they have the same dynamics (ordinal patterns probabilities are the same for both). Yet, the encoding of the external signal deteriorates when we change from the unidirectional to the bidirectional case [compare Fig. \ref{T1_2}(a) with Fig. \ref{Eel}(a) and Fig. \ref{T1_2}(c) with Fig. \ref{Eel}(b)]. This fact, underlies the difference between both types of neurons (class I and class II) when coupled electrically: while both keep the synchronization, the encoding of the signal is just kept for class I neurons.

\begin{figure}[!ht]
\centering
\includegraphics[width=0.5\columnwidth]{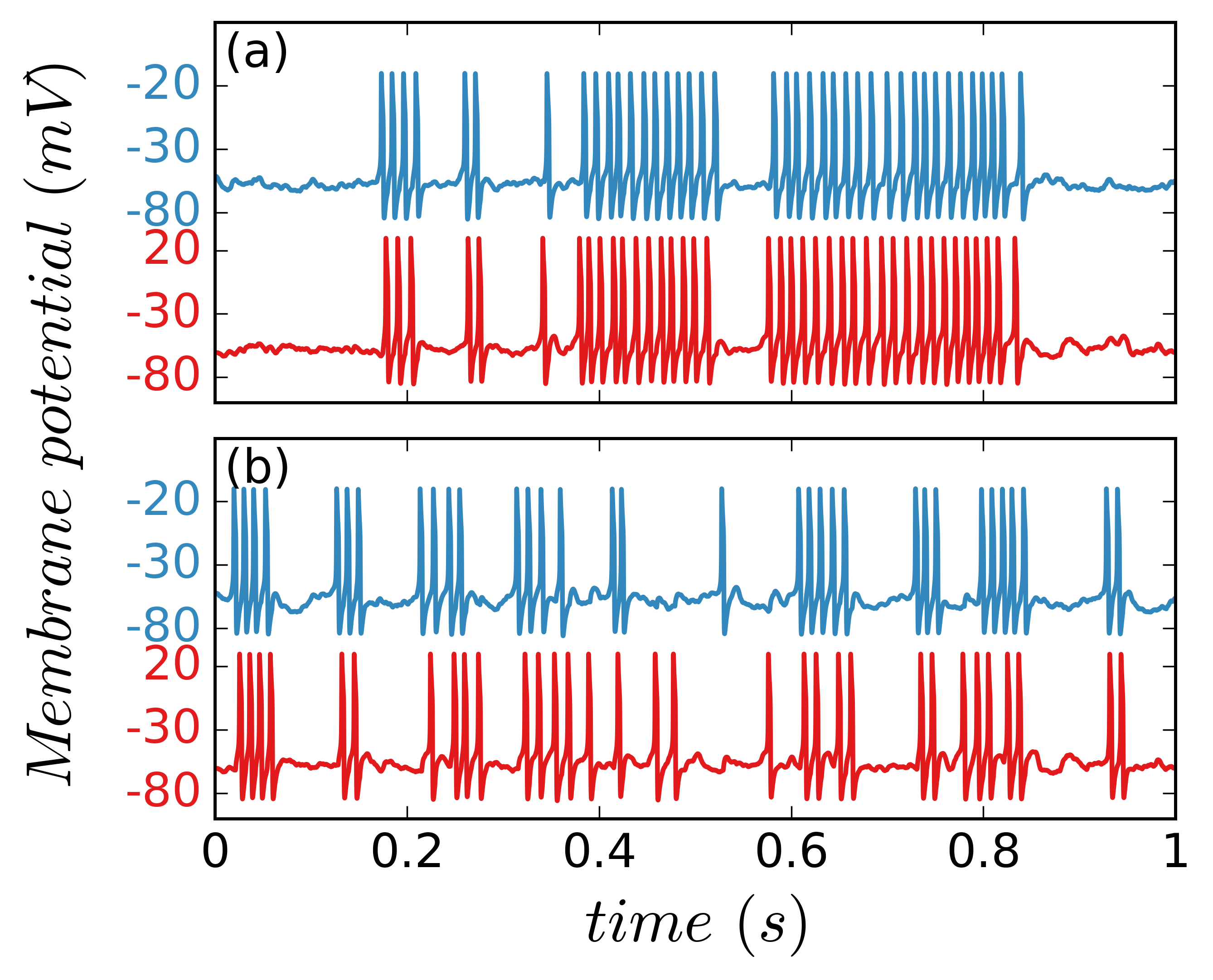}
\caption{Dynamics of neuron 1 (blue, top) and neuron 2 (red, bottom) when (b) the best encoding is achieved when $A_0 = 250$ mVcm$^2$/mS (neuron 1, class I and chemical bidirectional coupling; it corresponds to the observed maxima of 012 probability in Fig. \ref{E2} for $f = 10$ Hz). For comparison, panel (a) displays the neurons dynamics for the same parameters as panel (b) without modulation.}
\label{time-series-encoding}
\end{figure}

\begin{figure}[!ht]
\centering
\includegraphics[width=0.6\columnwidth]{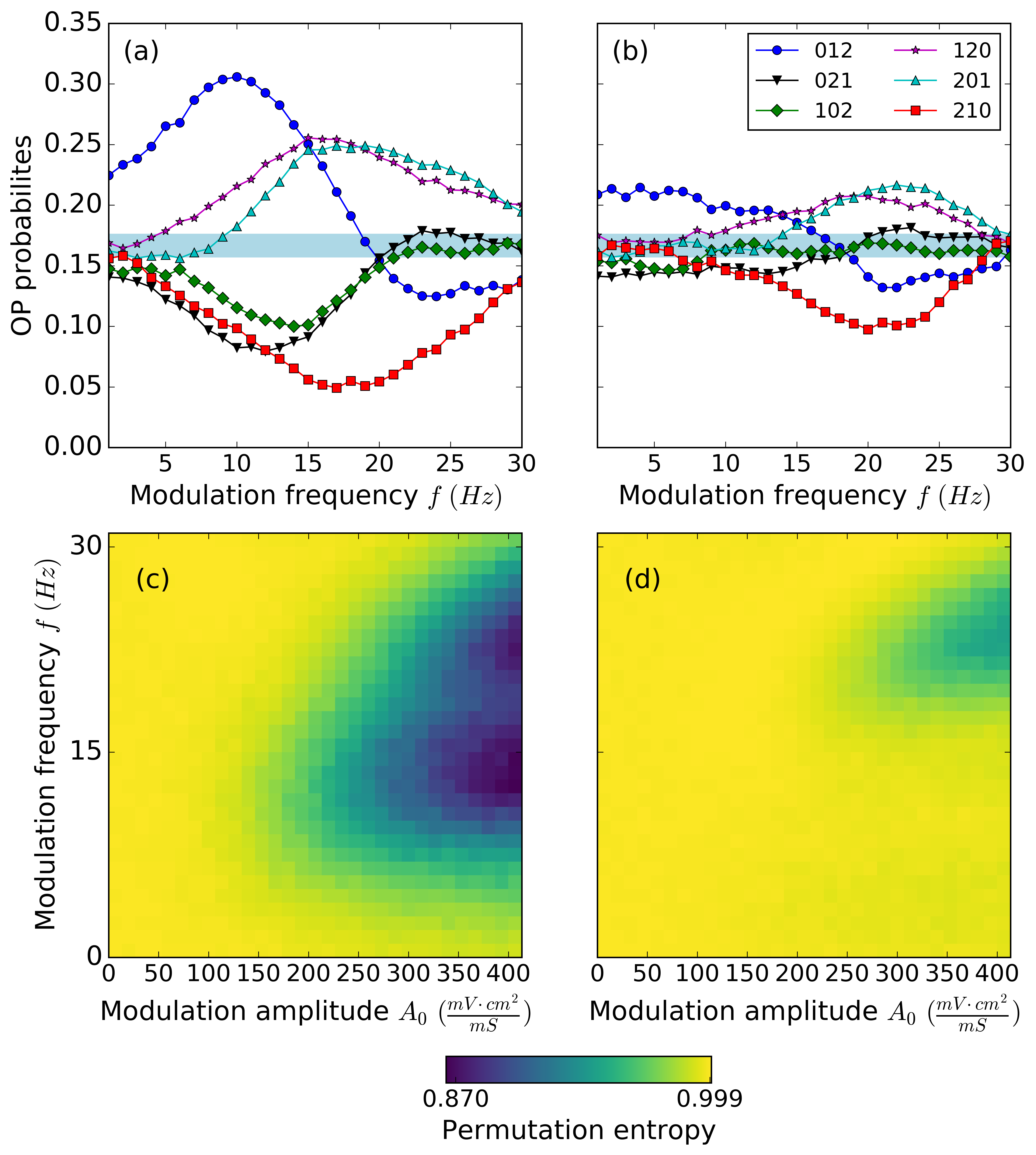}
\caption{OP probabilities as a function of the modulation frequency for neuron 1 (panel a) and neuron 2 (panel b), for $A_{0} = 250$ mVcm$^2$/mS. Permutation entropy, in color map, as a function of the frequency and the normalized amplitude of the external modulation for neuron 1 (panel c)  and neuron 2 (panel d). All panels correspond to class I and to bidirectional chemical coupling with $g_{{AMPA}_{norm}}$. For specific values see section \ref{param_values}.}
\label{E2}
\end{figure}

\begin{figure}[!ht]
\centering
\includegraphics[width=0.6\columnwidth]{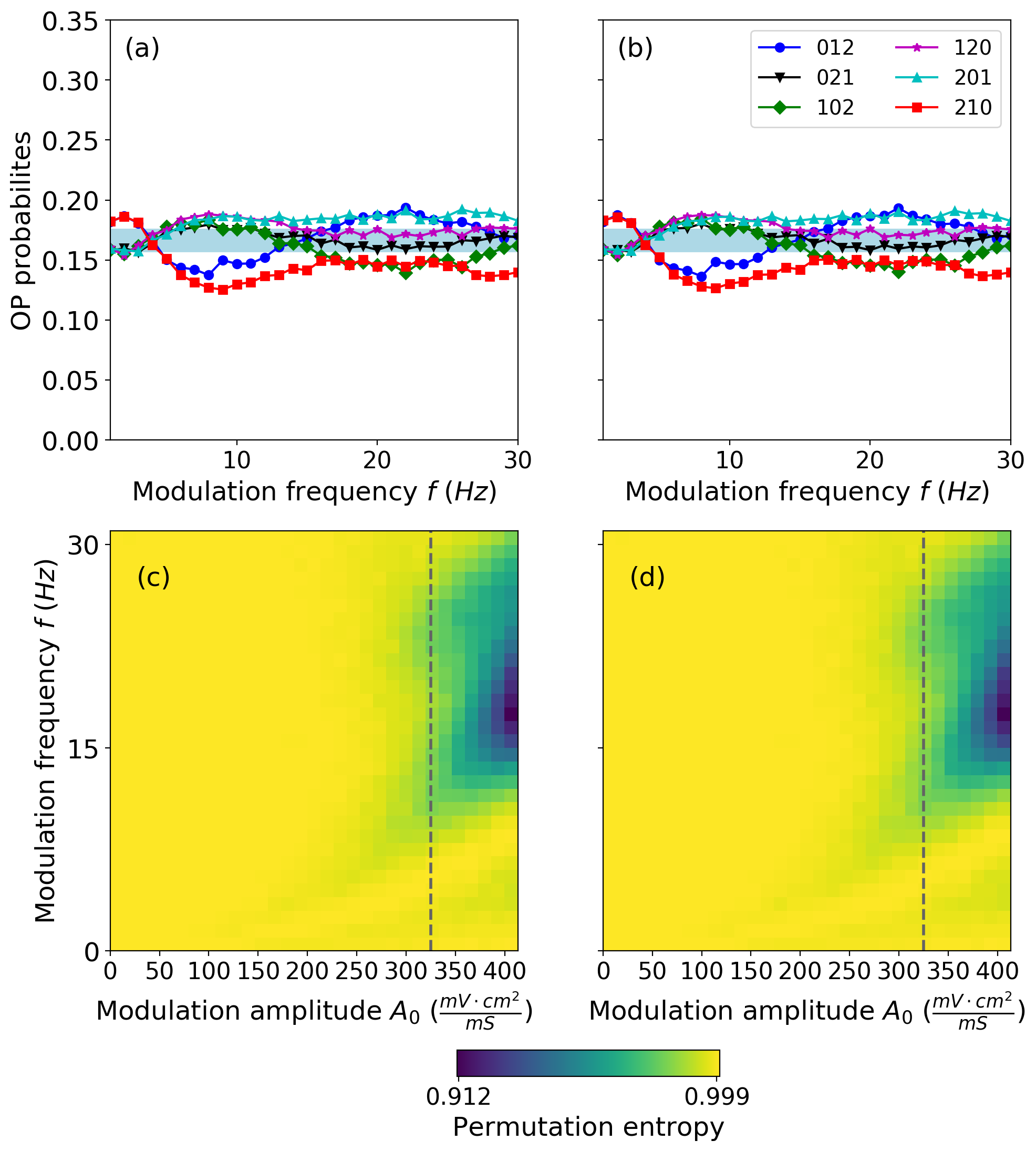}
\caption{OP probabilities as a function of the modulation frequency, for neuron 1 (panel a) and neuron 2 (panel b), for $A_{0} = 250$ mVcm$^2$/mS. Permutation entropy, in color map, as a function of the frequency and the normalized amplitude of the external modulation for neuron 1 (panel c)  and neuron 2 (panel d). All panels correspond to class II and to bidirectional electrical coupling with $g_{{gap}_{norm}}$. For specific values see section \ref{param_values}.}
\label{Eel}
\end{figure}

\begin{figure}[!ht]
\centering
\includegraphics[width=0.6\columnwidth]{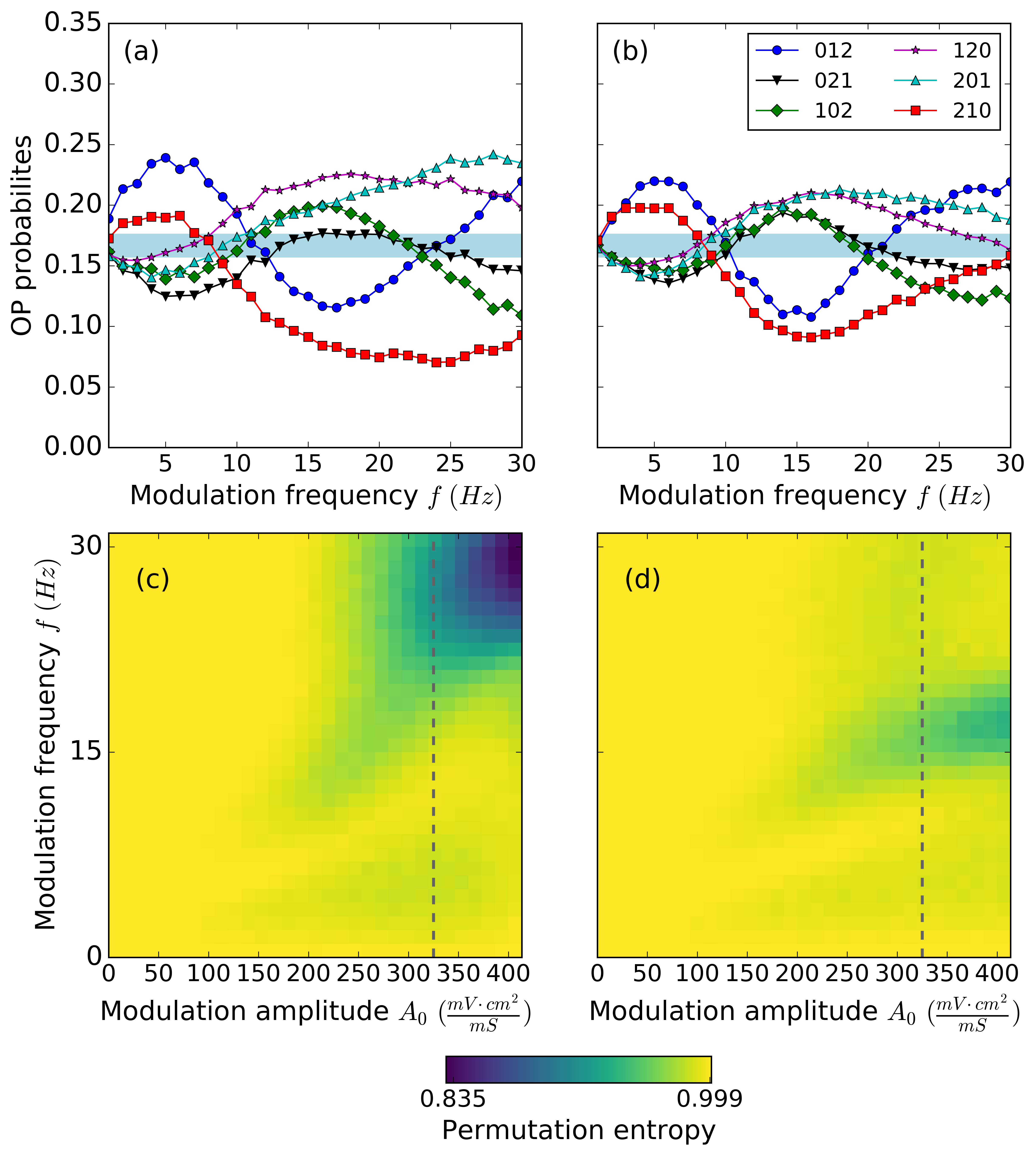}
\caption{OP probabilities as a function of the modulation frequency for neuron 1 (panel a) and neuron 2 (panel b), for $A_{0} = 250$ mVcm$^2$/mS. Permutation entropy, in color map, as a function of the frequency and the normalized amplitude of the external modulation for neuron 1 (panel c)  and neuron 2 (panel d). Dotted grey line marks when the modulation becomes supra-threshold ($A_0 \geqslant 325$ mVcm$^2$/mS). All panels correspond to class II and to bidirectional chemical coupling with $g_{{AMPA}_{norm}}$. For specific values see section \ref{param_values}.}
\label{E3}
\end{figure}

On the contrary, for the mutual chemical coupling there is a significant effect when compared with the unidirectional case. 
It can be seen in the permutation entropy [Figure \ref{E2} and \ref{E3}, panels (c) and (d)],  that the modulated neuron has a broader range of frequencies where information can be encoded.
For excitable class I neurons, the feedback from the second neuron strongly affects the modulated one improving the quality of the encoding and yielding small values of the permutation entropy at expenses of a reduced efficacy. 
This is due to the onset of the dominant pattern 012 for most of the frequencies that we have considered. 
In Fig. \ref{time-series-encoding}(b) we display the spiking dynamics of neuron 1 (blue, top) and neuron 2 (red, bottom) for this particular case: when the best encoding is achieved for $A_0 = 250$ mV~cm$^2$/mS) (that is for class I neuron with chemical bidirectional coupling for $f = 10$ Hz). 
For comparison in Fig. \ref{time-series-encoding}(a) we plot the dynamics for both neurons for the same regime as in panel (a) but without modulation. 

For class II neurons (Figure \ref{E3}), the modulated neuron has a similar efficacy as compared with the unidirectional coupling case (the neuron can identify two ranges of frequencies) although with a worse quality for the amplitude that we have considered ($A_{0}=250$ mV~cm$^2$/mS). 
In general, the small values of the permutation entropy shift to larger modulation amplitudes, which implies that, to be encoded, a stronger modulation signal is required.
It is worth mentioning that if the coupling strength in the chemical coupling is increased, the darker region in Figure \ref{E3}(c) is larger and its size approached to that observed in Figure \ref{E2}(c).

When two neurons are mutually coupled, independently of the excitability class, the receiver neuron changes its permutation entropy when compared with the unidirectional case. 
It has a larger entropy for most of the frequencies and modulation amplitudes, having just a small region of ordered spiking patterns [Figure \ref{E2}(b) and \ref{E3}(b)].
However, the changes of the receiver neuron allows the modulated one to change the way to encode the same information as compared with the unidirectional case, changing the efficacy and the quality of the encoding.

\begin{figure}[!ht]
\centering
\includegraphics[width=0.4\columnwidth]{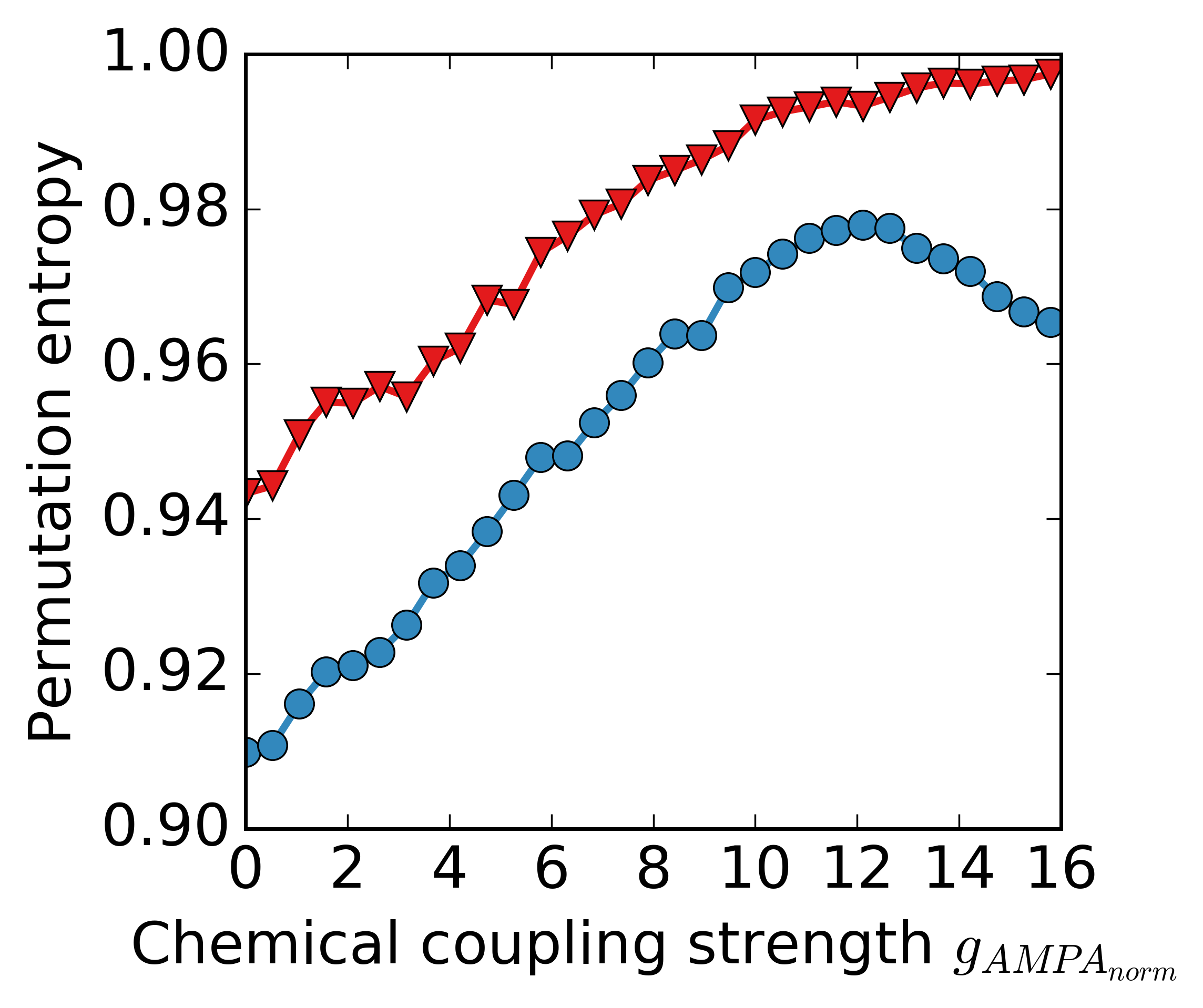}
\caption{Permutation entropy as a function of the chemical coupling strength (bidirectional) for neuron 1 (blue) and neuron 2 (red) for class II. Parameters:  $f=23$ Hz and $A_{0}=250$ mVcm$^2$/mS.}
\label{S1}
\end{figure}

The difference in the unmodulated neuron when comparing the unidirectional and bidirectional cases, is due to the coupling of the receiver neuron with the modulated one. For increasing values of the weights of the mutual coupling, we observe that the second neuron reaches a Plato at the maximum entropy whereas the permutation entropy of the modulated neuron decreases for a more significant conductance weight (Figure  \ref{S1}).
Thus, for the information coding, we can consider the bidirectional coupling as a single system and not as two individual components connected between them. This means a system with a particular complexity in its dynamics that allows to encode the information differently and provides different features than an isolated neuron.
	
\subsection{\label{Transmission}Information transmission in the presence of different couplings}
\begin{figure}[!ht]
\centering
\includegraphics[width=0.5\columnwidth]{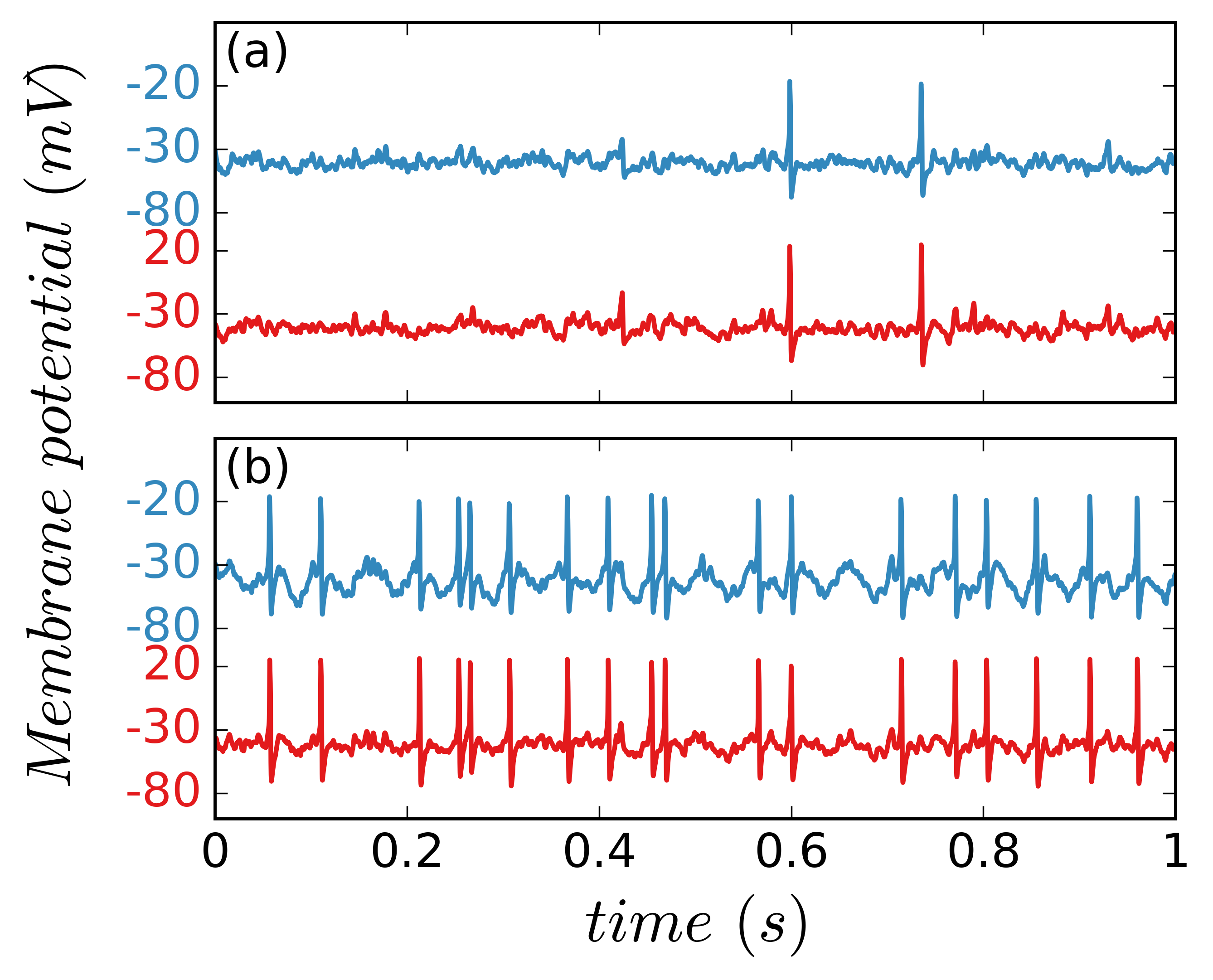}
\caption{Dynamics of neuron 1 (blue, top) and neuron 2 (red, bottom) when (b) the best transmission is achieved when $A_0 = 250$ mV~cm$^2$/mS (neuron 2, class II and electrical unidirectional coupling; it corresponds to the observed minima of 210 probability in Fig.  \ref{T1_2}(e), for $f = 20$ Hz). For comparison, panel (a) displays the neurons dynamics for the same parameters as panel (b) without modulation.}
\label{time-series-trans}
\end{figure}

Information transmission depends on the characteristics of the transmitter and the receptor. 
For neurons described by a single compartment model, the transmission of information between two of these neurons could be considered when the post-synaptic neuron copies, in a certain way, the behavior of the pre-synaptic one. 
Following this idea, we study the influence of the excitability class and type of synapses of the neurons in the unidirectional and bidirectional coupling scheme (Figure \ref{Scheme}). 
In the brain, the chemical synapses between pre- and post-synaptic neurons are more abundant than the electrical ones (also called gap junction).
Also the former can be unidirectional or bidirectional, while the latter are mostly bidirectional although in certain cases a preferred direction for the information flow is established. 
In any case, both kinds of synapsis contribute in a way or another to the communication in the brain.
The effect of each linkage in the information transmission turns out to be independent of the excitability class of the neurons. 
The electrical coupling, for instance, reaches the best performance in both class I and class II neurons since the receiver has the same permutation entropy, as seen, for the unidirectional coupling, in Figure \ref{T1_1}(f) and \ref{T1_2}(f), giving equal probabilities of the OP for a given amplitude, represented in Figure  \ref{T1_1}(e) and \ref{T1_2}(e), as compared to the modulated neuron. 
This result is expected due to the diffusive nature of the gap junction that tends to synchronize the two neurons. 
Fig. \ref{time-series-trans}(b) displays the spiking dynamics of neuron 1 (blue, top) and neuron 2 (red, bottom) for the case where the best transmission is achieved  (fixing $A_0$ to 250 mV~cm$^2$/mS) for class II neurons for a frequency of the external signal $f = 20$ Hz  (it corresponds to the minima of 210 probability in Fig. \ref{T1_2}(e)). For comparison Fig. \ref{time-series-encoding}(a) shows the dynamics for both neurons for the same regime as in panel (a) but without modulation.

Under chemical synapses the results are different.
In the unidirectional model and excitatory chemical synapse, there is an attenuation of the transmitted information as can be seen in Figures \ref{T1_1}(d) and \ref{T1_2}(d), where the entropy at the receiver neuron is higher than that of the modulated one. 
Therefore OP probability differences are smaller, as shown in Figures \ref{T1_1}(c) and \ref{T1_2}(c). 
In this case, the noise influences more than in the gap junction case where the synchronization is favored.
Nevertheless, unlike the gap junction, the chemical synapse enriches the system's response in the presence of a bidirectional coupling (Figures \ref{E2} and \ref{E3}). 
In this case, both neurons communicate each other but, in terms of the information defined in this study, the signal cannot be considered as transmitted for most of its frequencies and amplitudes. 
This is due to the change of the permutation entropy that drastically increases for the receiver neuron.
This effect contributes to the idea that two mutually-coupled neurons can be considered as a single particular system, where the neuron that receives the external signal can encode information in a particular way due to the presence of the second neuron.

\section{\label{conclu} Conclusions}
We have studied the neuronal encoding and transmission of a weak periodic external signal using a modification of the Morris-Lecar model, which allows to vary the neuron's excitability class. We have considered two neurons, one that perceives the weak signal and another that does not perceive it. The two neurons interact through different types of coupling (unidirectional or bidirectional, chemical or electrical). To quantify the encoding and the transmission of the signal we have applied symbolic ordinal time series analysis to the sequences of inter-spike-intervals of each neuron. Analyzing the probabilities of symbolic spike patterns, and the permutation entropy computed from the symbolic probabilities, we have studied how information encoding and transmission depend on the excitability class of the neurons, and of the type of connection.

With unidirectional coupling, we have found that the neuron that perceives the signal is more or less sensitive to the frequency of the signal, depending on its excitability class. Class I neurons express patterns that are better resolved at low frequencies, whereas class II neurons have a better differentiation at higher frequencies [see Figs. \ref{T1_1}(b) and \ref{T1_2}(b)]. 

When the neurons are class I, the synaptic mutual coupling can significantly improve the encoding of the signal. 
As it can be seen by comparing Figs. \ref{E2}(c) and \ref{T1_1}(b), the chemical coupling increases the range of frequencies that can be encoded. 
Instead, the electric coupling has no significant effect in the signal encoding [compare Figs. \ref{E1}(c) and \ref{T1_1}(b)]. 
In class II neurons, chemical coupling (bidirectional and unidirectional) does not affect the encoding of the signal whereas for the electrical coupling (bidirectional) the encoding of the sub-threshold signal deteriorates [compare Figs. \ref{E3}(c) and \ref{T1_2}(b)]. 

Regarding the transmission of the signal, for both excitability classes and connectivity models, electric coupling is the best mechanisms to transmit the information, as the second neuron expresses the symbolic patterns with nearly the same probabilities as the neuron that perceives the signal. 
This is because electrical coupling tends to synchronize both neurons. 
Therefore, the transmission is more efficient due to the diffusive effect from the first neuron to the second one.

In the case of chemical coupling, when the neurons are unidirectionally coupled the information is transmitted only if the neurons are of class II [Figs. \ref{T1_2}(d) and Figs. \ref{T1_2}(b)]; in contrast for class I neurons unidirectional synaptic coupling does not allow the information to be transmitted [Figs. \ref{T1_1}(d) and Fig. \ref{T1_1}(b)].
For the bidirectional synaptic coupling, the receiver neuron does not express the same patterns as those expressed by the neuron that perceives the signal [for both, class I and class II excitability, as seen in Figs. \ref{E2}(d) and \ref{E3}(d)]. 
Therefore, the information is not properly transmitted. 
Nevertheless, for class I excitability the two coupled neurons can be considered as a single unit that can codify information in a wider range of signal frequencies, as compared to the individual neuron [Figs \ref{E2}(c) and \ref{T1_1}(b)].

 In general, the bidirectional coupling improves the capacity of encoding the signal information. Class I neurons are more susceptible to the two-way coupling, changing the expressed patterns when compared with the unidirectional case. Information transmission is very high in the presence of the electrical coupling due to its diffusive properties. 
 On the contrary, for the case of chemical bidirectional coupling there are changes in the dynamics of both neurons, potentially providing a new way of encoding features.

\newpage
\appendix
\section{\label{Appendix}Excitability of the neuron}
 
\begin{figure}[!ht]
\centering
\includegraphics[width=0.6\columnwidth]{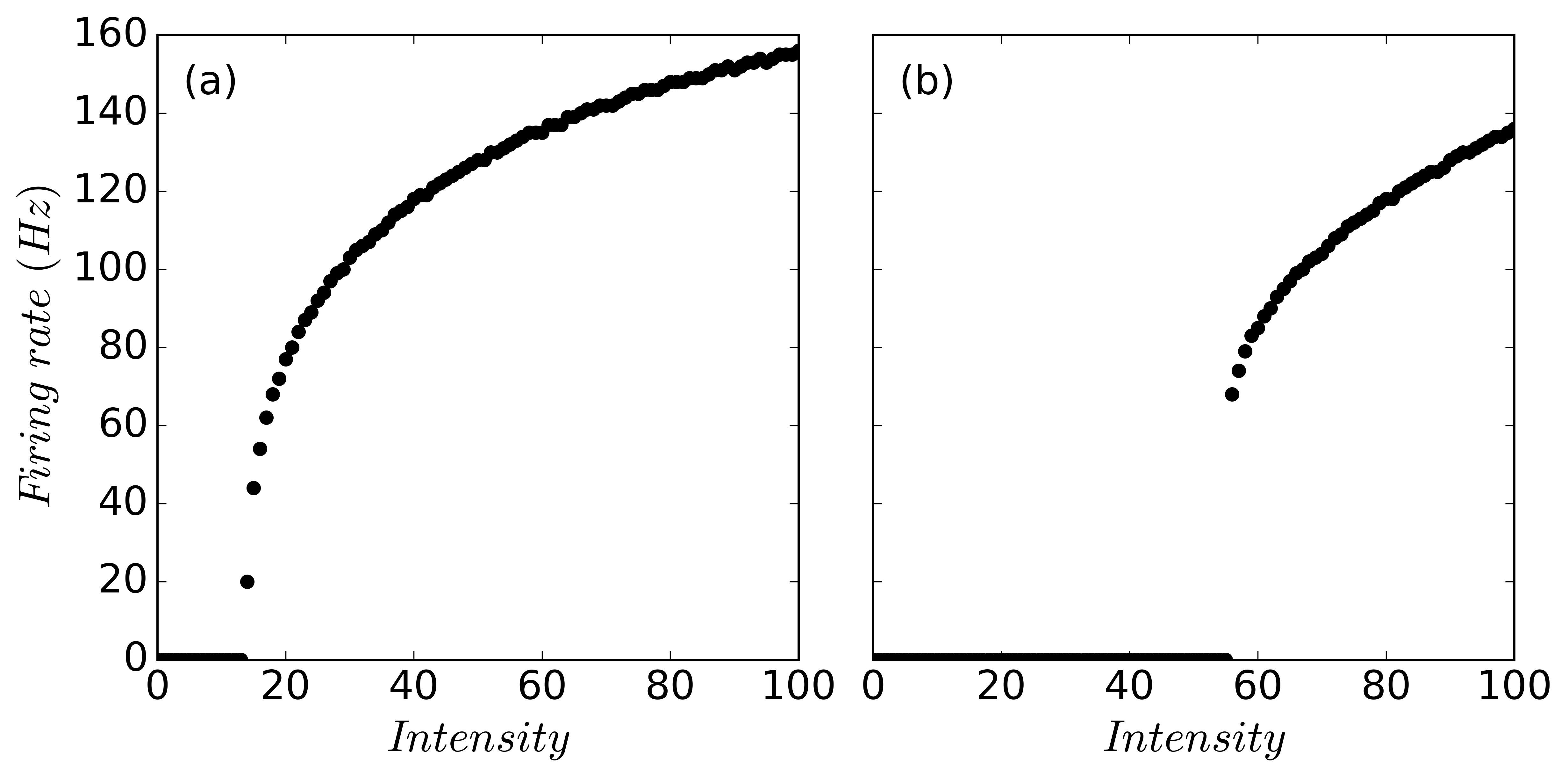}
\caption{Current-frequency curve for (a) class I and (b) class II with $\beta_m=-12$ mV.}
\label{c1}
\end{figure}

Table \ref{param} shows the value used for each parameter for each class of neuron. Class 1 responses at zero frequency and the firing rate increases gradually [Figure \ref{c1}(a)], whereas the response of Class 2 emerges at non zero frequency [Figure \ref{c1}(b)].  
 
\begin{table}[ht!]
\centering
\begin{tabular}{|l|c|c|}
\hline
Parameters & Class I & Class II\\
\hline
$E_{Na}(mV)$ & 50 & 50\\
\hline
$E_{K}(mV)$ & -100 & -100\\
\hline
$E_{leak}(mV)$ & -70 & -70\\
\hline
$g_{fast}(mS/cm^2)$ & 20 & 20\\
\hline
$g_{slow}(mS/cm^2)$ & 20 & 20\\
\hline
$g_{leak}(mS/cm^2)$ & 2 & 2\\
\hline
$C(\mu F/cm^2)$ & 2 & 2\\
\hline
$\phi_{w}(mS/cm^2)$ & 0.15 & 0.15\\
\hline
$\beta_{w}(mV)$ & -10 & -10\\
\hline
$\gamma_{w}(mV)$ & 13 & 13\\
\hline
$\gamma_{m}(mV)$ & 18 & 18\\
\hline
$\beta_{m}(mV)$ & -12 & 0\\
\hline
\end{tabular}
\caption{Parameters of the Morris-Lecar model used for the simulation of the neuron class I and II.}
\label{param}
\end{table}

\begin{acknowledgments}
CE and CRM acknowledge the Spanish State Research Agency, through the Mar\'{\i}a de Maeztu Program for Units of Excellence in R \& D (MDM-2017-0711). CE is funded by the Conselleria d'Innovaci\'o, Recerca i Turisme of the Government of the Balearic Islands and the European Social Fund  with grant code FPI/1900/2016. MM and CM acknowledge support from the Spanish MINECO/FEDER grant FIS2015-66503-C3-2-P141 and ICREA ACADEMIA, Generalitat de Catalunya.
\end{acknowledgments}


\end{document}